# Giant electrostriction in bulk RE (III) substituted CeO$_2$: effect of RE$^{3+}$ and its concentration.

Soumyajyoti Mondal [a], Pooja Punetha [a,b], Rajeev Ranjan [b], Pavan Nukala* [a].

[a] Center for Nano Science and Engineering, Indian Institute of Science, Bengaluru-560012, India.

[b] Materials Engineering, Indian Institute of Science, Bengaluru, 560012- India.

Corresponding author (*) email: pnukala@iisc.ac.in

**Abstract** – Recent discovery of giant electrostriction in rare earth (RE (III)) substituted ceria (CeO$_2$) thin films driven by electroactive defect complexes and their coordinated elastic response, expands the material spectrum for electrostrain applications beyond the conventional piezoelectric materials. Especially Gd substituted CeO$_2$, with Gd concentration >10% seems to be an ideal material to obtain such large electrostrain response. However, there are not many experimental studies that systematically investigate the effect of RE (III) ion-defect interaction and RE concentration on electrostriction. Here we perform structure-property correlation studies in bulk ceramics of RE$^{3+}$ substituted ceria doped with RE=Y, La and Gd at various concentrations upto a maximum of 20%, to understand the features responsible for giant electrostriction. Our results show that Y substituted ceria, with atleast 20% Y substitution, is clearly both a giant M and a Q electrostrictor at low frequencies (<20 Hz), and this correlates with the unique attractive defect-dopant interaction of Y with $V_O^{\bullet\bullet}$. La has a repulsive interaction with $V_O^{\bullet\bullet}$, and La doped ceria at all the studied compositions (upto 20%) does not show giant electrostiction. Gd has a neutral interaction, and only 20% Gd doped ceria at best falls at the border of classification between giant and non-giant electrostrictors at frequencies <0.05 Hz. Our work takes a step back from thin-films and assesses the fundamental defect features required in the design of giant electrostrictors.

**Keywords:** Giant electrostriction, vacancies, point defects, electroceramics, X-ray diffraction.

**Introduction:**

Materials with large electromechanical response are crucial for applications in nano/micro electromechanical systems (N/MEMS) such as sensors, actuators, transducers, nanopositioners and so on[1,2]. Traditionally, large piezoelectric coefficients of non-centrosymmetric Pb-based materials such as $Pb(Zr,Ti)O_3$ (PZT), and large electrostrictive coefficients of centrosymmetric materials such as $0.9[Pb(Mg_{1/3}Nb_{2/3})O_3]$-$0.1[PbTiO_3]$ (0.9PMN-0.1PT) have been exploited for these applications [3–5]. However, the quest to go Pb-free has triggered immense research in other piezoelectric materials such as $K_{0.5}Na_{0.5}NbO_3$-$xBaTiO_3$ (KNN-BT), aimed at designing flat thermodynamic energy profiles that ease the polarization rotation[6,7]. In the realm of electrostriction, a second order electromechanical effect, 'giant' electrostriction was discovered in 20% Gd-substituted $CeO_2$ thin film in 2012[8]. It has by now been established that Gd substitution provides oxygen vacancies which interact with dopants creating electroactive defect complexes. A coordinated elastic response from these complexes upon the application of an electric field generates a large second order electrostrain response[9]. Giant electrostriction has also been reported in Gd-substituted $CeO_2$ thin films with other concentrations of Gd >10% [10].

2$^{nd}$ order electrostrain can be expressed by $x = QP^2$ or $x = ME^2$, where $x$, Q, P, M, and E are electrostrain, polarization electrostriction coefficient, polarization, field electrostriction coefficient, and electric field respectively. In classical electrostrictors Q (hydrostatic) and M can be empirically expressed in terms of s (elastic compliance) and ε (dielectric permittivity) as follows:

$$|Q_h| \approx 2.3 \left(\frac{s}{\varepsilon_0 \, \varepsilon_r}\right)^{0.59} \quad \text{(Newnham's relation)} \quad \text{(eq. 1)[11]}$$

$$|M| \approx 10^4 (s\varepsilon_0 \, \varepsilon_r)^{1.14} \quad \text{(Yu and Janolin's relation) (eq. 2) [12]}$$

As per Yu and Janolin [12], materials that exhibit Q value, one order of magnitude larger than equation 1, are classified as 'giant' Q electrostrictors. Similarly, materials that exhibit M value, one order of magnitude larger than equation 2, are classified as 'giant' M electrostrictors, and it is in general not necessary that giant M electrostrictors are also giant Q electrostrictors [13]. In this work, we utilize the above definitions of giant electrostrictors to classify various $RE^{3+}$ substituted $CeO_2$ as a function of $RE^{3+}$ concentration (5 to 20%). La, Gd and Y were chosen as the $RE^{3+}$ ions, all of them larger than $Ce^{4+}$ ions. The criterion for selecting these dopants arises from a DFT based study by Nakayama et al., who showed that the defect association energy of $RE^{3+}$ ion with $V_O^{\bullet\bullet}$ is repulsive with La, neutral with Gd and attractive with Y [14]. These systematic studies were carried out on bulk ceramic samples, to avoid any extrinsic effects arising from surfaces and residual stresses that are common in thin films.

**Experimental:**

RE substituted $CeO_2$ bulk ceramic samples were synthesized by solid state reaction method. $CeO_2$ and $RE_2O_3$ (RE = Gd, La, Y) powders were mixed in stoichiometric ratio and ball milled in acetone medium using zirconia vials and balls at 200 rpm for 12 hours. Ball milled powders were dried, ground and calcined at 1450°C for 10 hours in alumina crucibles. X-ray powder diffraction was performed using Rigaku SmartLab diffractometer with monochromatic CuK$_\alpha$ radiation. 5% PVA binder is mixed with calcined powders and pressed into pellets using 10 mm diameter die with uniaxial pressure and followed by cold isostatic pressure. Pellets were sintered in alumina crucibles at 1500°C for 10 to 50 hours, until a density >95% (estimated through Archimedes principle) was achieved. Pellets were thinned down to 0.8 mm and annealed at 1480°C to remove residual stresses (see Figure S1 for SEM images of the pellets). Silver paste was applied to make electrical contacts. Electrostrain was measured using Precision Premier II Ferroelectric Tester with Trek amplifier and MTI-2100 Fotonic™ Sensor.

**Results and Discussion**:

Figure 1 compares (220) Bragg peak positions of $RE_xCe_{1-x}O_2$ at various RE concentrations (x). From here on we will also refer to these compositions as y RE SC, indicating y% RE substituted Ceria (y=100 x). Note that for all the selected RE ions, Y, La and Gd, the ionic size of RE ion is larger than that of $Ce^{4+}$, that they substitute [14]. In the cases when RE=Gd and La (Figure 1a and b respectively), the (220) peaks (and all the other peaks) shift to a lower angle with increasing x, consistent with larger size of $RE^{3+}$ cation in comparison to $Ce^{4+}$. Surprisingly, when RE=Y (Figure 1c), all the peaks shift to a larger angle indicating a lattice contraction despite $Y^{3+}$ having a larger ionic radius than $Ce^{4+}$. Lattice parameters in each case were estimated from Rietveld refinement of full powder XRD patterns (Figure S2), and consolidated in Figure 1d. The slope of increase in lattice parameter with x is larger when RE=La compared to when RE=Gd, and is negative (decreases) when RE=Y. Such anomalous trends in lattice parameter are correlated with the defect-dopant ($RE^{3+}$) interactions reported by Nakayama et al., in ref [14]. For La substituted Ceria (LSC) both the larger size of $La^{3+}$ and its repulsive interaction with $V_O^{\bullet\bullet}$ contribute towards a steeper increase in lattice parameter with increasing x than Gd substituted ceria (GSC), where Gd has a neutral interaction with $V_O^{\bullet\bullet}$. In case of $Y^{3+}$ substituted ceria, however, the attractive defect-dopant interaction contracts the lattice, dominating over the expansion effects resulting from the larger ionic size of $Y^{3+}$.

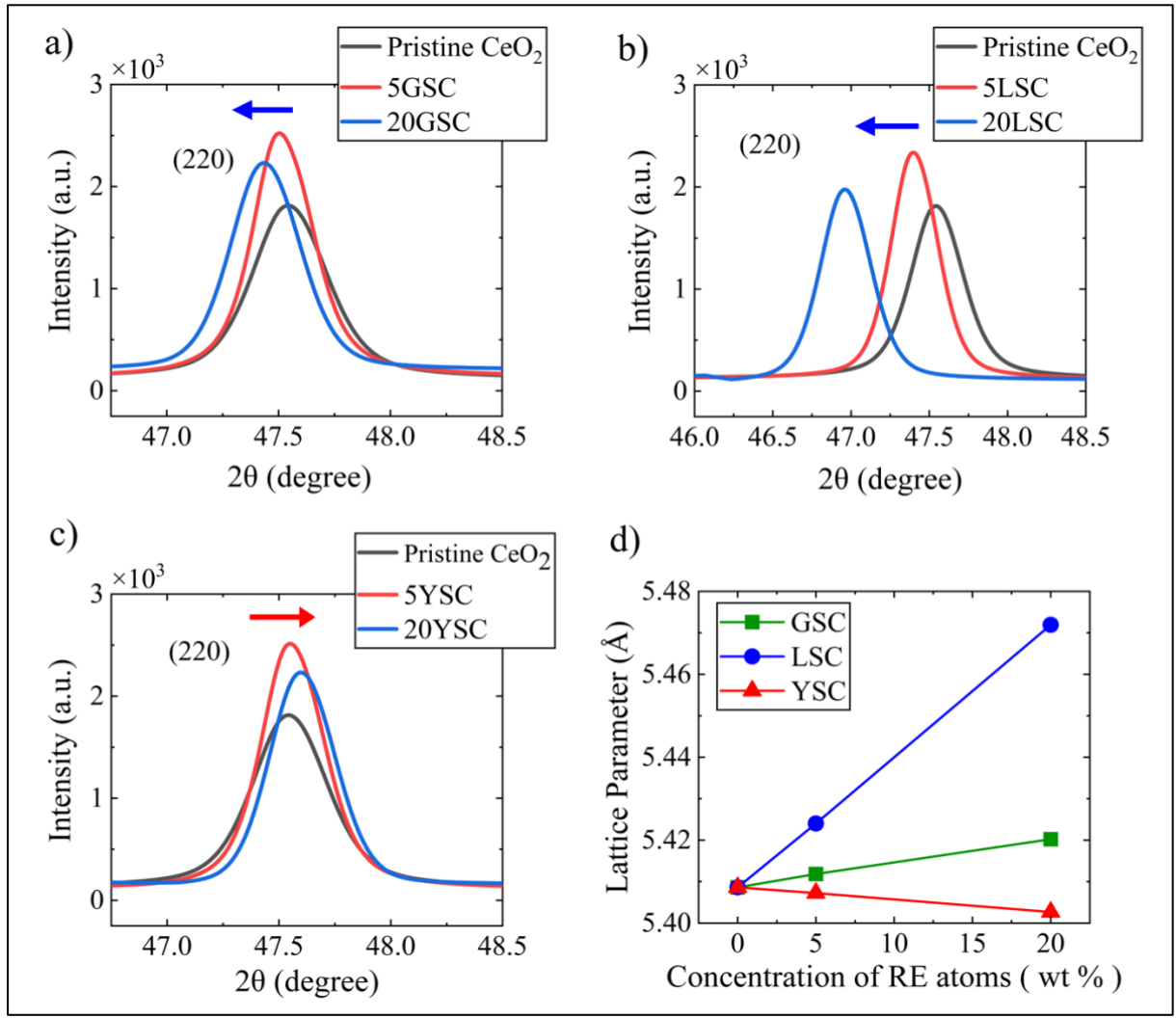

**Figure 1. Comparison of (220) Bragg peak of $\theta$-$2\theta$ powder XRD patterns** a) Pristine $CeO_2$, 5GSC and 20GSC; b) Pristine $CeO_2$, 5LSC and 20LSC; c) Pristine $CeO_2$, 5YSC and 20YSC; d) Change in lattice parameter due to RE atom substitution.

To understand the structure-defect-electrostriction correlation, we performed electrostrain measurements at various frequencies (0.03 Hz to 10 Hz) on all the samples discussed above. In Figure 2, we compare electrostrain values with a systematic variation of dopant type (Figure 2a), dopant concentration (Figure 2b) and frequency (Figure 2c). In compositions of 20YSC, we observe significant electrostrain of 0.03% at 100 kV/cm and 0.1 Hz. The strain-field loops show butterflylike hysteresis indicative of underlying electromechanical dissipative processes. In comparison to 20YSC, when 20GSC and 20LSC, show smaller electrostrain values (~0.009% for Gd and 0.004% for La at 100 kV/cm and 0.1 Hz, Figure 2a). For YSC compositions studied, 20% Y substitution shows much larger electrostrain (~0.03% at 100 kV/cm and 0.1 Hz), compared to the other concentrations. Similar trends are observed with other dopants too (La and Gd). Finally, electrostrain increases with decrease in frequency and saturates at lower frequencies (<1 Hz, Figure 2c) which is a consequence of a defect induced phenomenon.

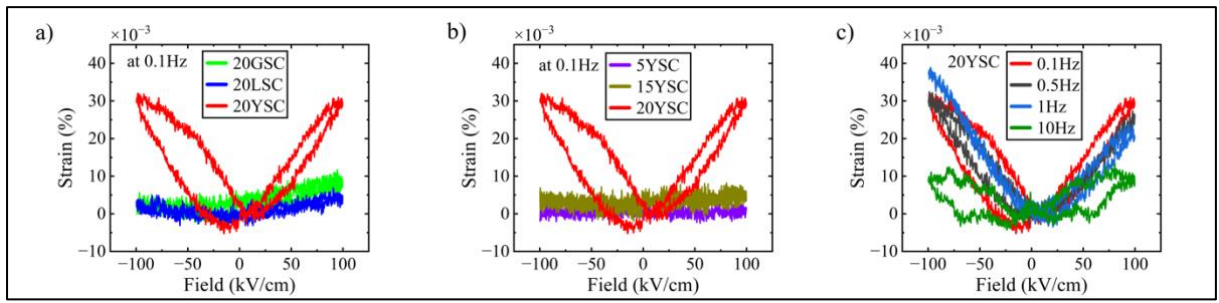

**Figure 2.** Electrostrain as a function of electric field a) for $RE_{0.2}Ce_{0.8}O_2$ (RE= Gd, La, Y), b) for $Y_xCe_{1-x}O_2$ (x= 5, 15, 20%) and c) for $Y_{0.2}Ce_{0.8}O_2$ at various frequencies.

The electrostrictive M coefficient ($M_{33}$) was estimated as the slope of strain ($x$) vs $E^2$ plots in all $RE_xCe_{1-x}O_2$ ceramics, and is consolidated in Figure 3. The noise floor for strain measurements is in the order of $10^{-5}$ at all the frequencies for all the samples. This puts a sensitivity limit on the measurement of $M_{33}$ to be ~$10^{-19}$ $m^2/V^2$, and so it must be noted that if the M for any sample is not reported in Figure 3, it is below the sensitivity limit.,. Following the definition of giant M electrostrictor given by Yu and Janolin, only those samples which are shaded in red in Figure 3 classify as giant electrostrictors [12]. Clearly, 20YSC samples fall into this category below 20 Hz, and 20GSC samples below 0.05 Hz. Note that the frequency dispersion of $M_{33}$ in 20% Gd substituted $CeO_2$ is typical of defect-induced non-classical electrostriction, which however needs to be distinguished from a "giant" electrostrictor. In other words, non-classical electrostrictors always need not be giant electrostrictors.

From these results, we note that the following factors are crucial to design giant M electrostrictors in acceptor substituted ceria, which are solely driven by electroactive defects (such as in bulk systems):

a) An attractive vacancy-acceptor ion interaction as present in yttria substituted ceria

b) concentration of the acceptor atom of at least 20%, which creates enough electroactive defects that give a coherent strain response.

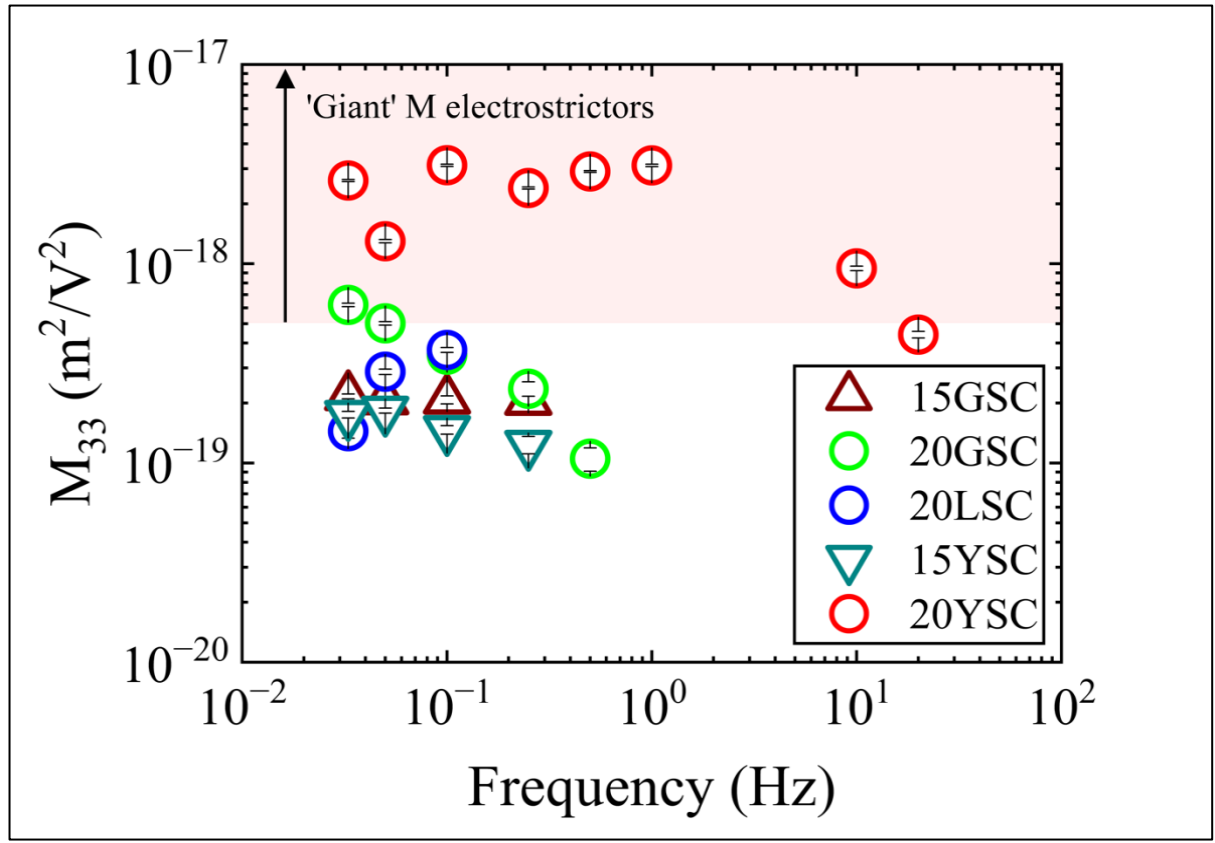

**Figure 3. Consolidation of $M_{33}$ values of $RE_xCe_{1-x}O_2$ as a function of frequency.**

Next, we evaluated the $Q_{33}$ coefficients of our samples to acquire necessary guidelines to design giant Q electrostrictors. To this end, we measured the induced polarization of all the samples synchronously with the electrostrain (see displacement (D) - electric field (E) loops in supplementary information, Figure S3). Note that in the limit of electric field we applied, displacement response seems linear with electric field.

The corresponding strain-displacement curves comparing between various acceptor ions (Figure 4a), various concentrations of an acceptor ion (Figure 4b) again show a dominant effect for 20YSC samples. The $Q_{33}$ estimated as the slope between $x$ and $D^2$ as a function of frequency is consolidated in Figure 5, on all the samples on which electrostrain was measurable. The giant Q electrostrictors are marked in the blue band, and only 20YSC samples belong to this category.

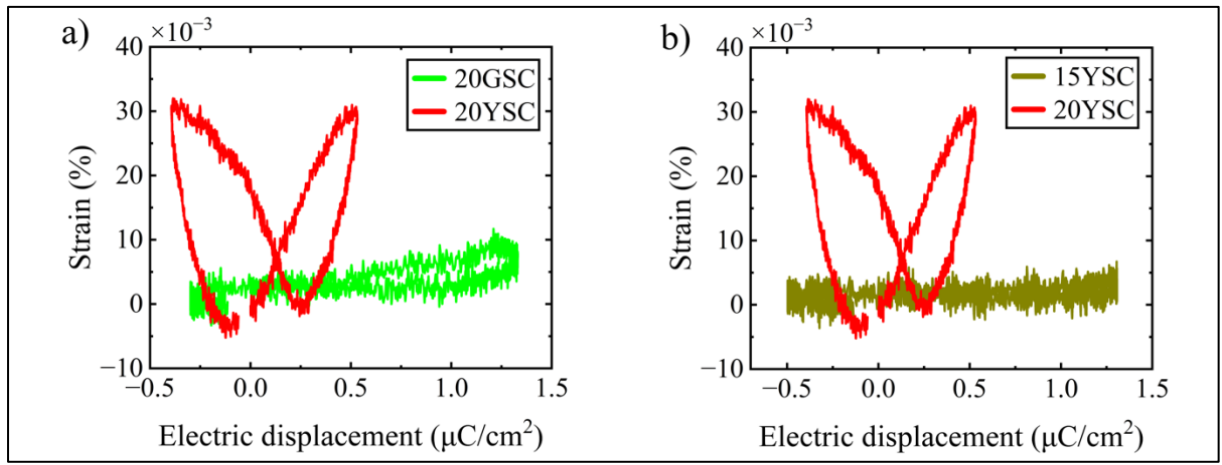

**Figure 4. Electrostrain as a function of electric displacement** a) for $RE_{0.2}Ce_{0.8}O_2$ (RE= Gd, Y), b) for $Y_xCe_{1-x}O_2$ (x=15, 20%).

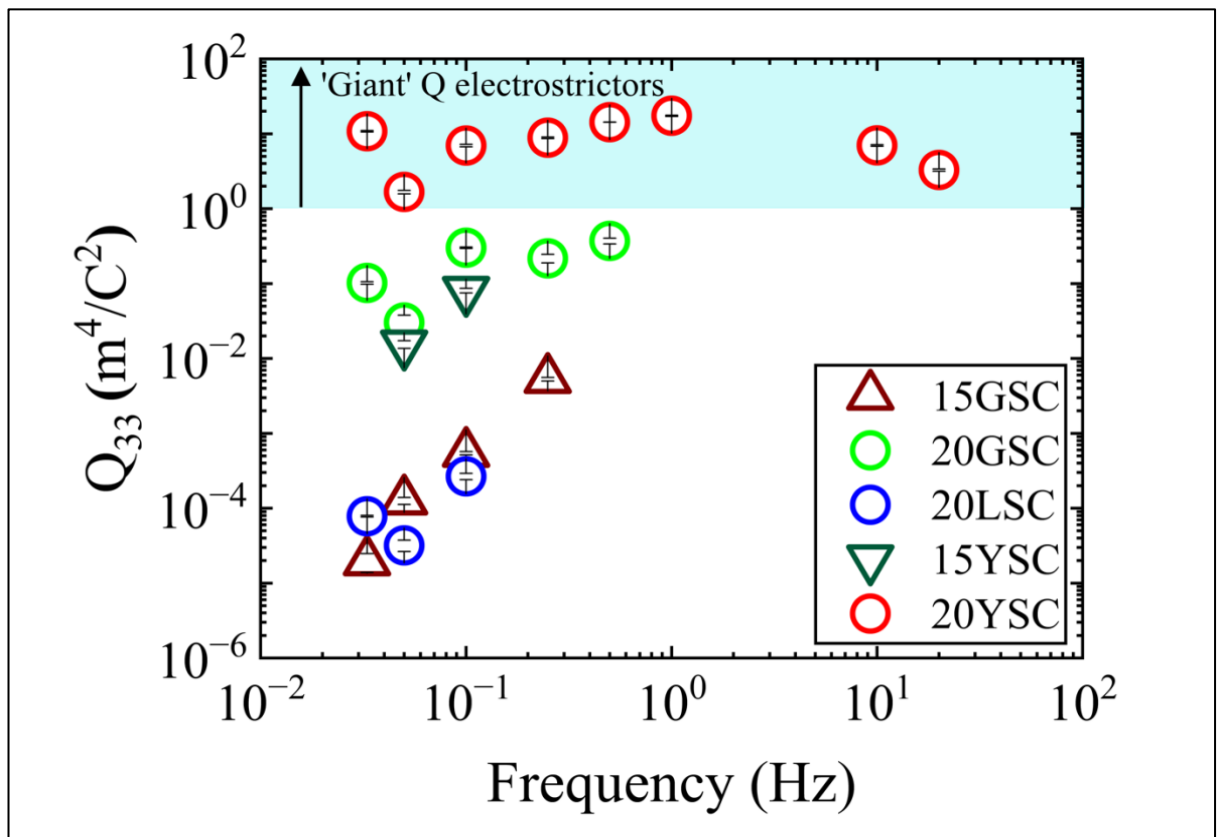

**Figure 5. Consolidation of $Q_{33}$ values of $RE_xCe_{1-x}O_2$ as a function of frequency.**

These results again clearly show that attractive interaction of acceptor ion with oxygen vacancy, in addition to at least 20% concentration is a recipe to engineer both giant M and Q electrostrictors.

Most of the works in literature [8,9,15,16] studied giant electrostriction on 20% Gd substituted ceria thin films. In these systems, the values of M and Q were reported to be larger than what we show here ($M_{33} \sim 10^{-18}$ to $10^{-16}$, and $Q_{33} \sim 10^1$ to $10^2$ m$^4$/C$^2$) [10]. Furthermore, in ref [10], the authors have reported larger M and Q coefficients 10GSC thin films than in 20GSC or 33GSC thin films. We propose that these enhanced effects can a result of not just electroactive defects, but also the other possible extrinsic effects that thin films possess such as residual stresses, microstructure, electrode clamping effects and so on, which need to be investigated. Here, we also mention another comparative study performed on bulk ceramics of acceptor doped ceria, which also reports larger M values for 10% or less Gd, La substituted ceria, than what we observe. We note that this report studies these materials only in the low field regime (E<10 kV/cm), which renders larger apparent slopes in $x$-$E^2$ plots as a consequence of hysteresis (as we also observe)[17].

**Conclusions:**

We report a systematic study on giant electrostriction in Re (III) substituted ceria, by comparing the effect of different acceptors which are larger than Ce (Y, Gd, La) and by varying their concentrations (5 to 20%). We show that in bulk, only 20YSC is both a giant M and Q electrostrictor as per the strict definitions given by Yu and Janolin. Different from reports on thin films, 20GSC is a giant M electrostrictor, only below 0.05 Hz, highlighting a case for studying the effect of other extrinsic factors that appear in thin film systems. We learn that an attractive interaction between acceptor ion and oxygen vacancy, in addition to a minimum concentration of the acceptor ions creates the maximum defect induced electrostrictive effects. The vacancy-acceptor interactions for acceptors larger in ionic size than ceria proposed in DFT study by Nakayama et al., is also verified from our XRD results. Our structure-property-defect correlation study lets us propose the use of 20YSC thin films to achieve much greater "giant" electrostrictive effects in devices.

**Acknowledgements:**

This work was partly carried out at Micro and Nano Characterization Facility (MNCF) located at CeNSE, IISc Bengaluru, funded by NPMAS-DRDO and MCIT, MeitY, Government of India; and benefitted from all the help and support from the staff. We also acknowledge support received from Prof. Rajeev Ranjan's lab, Materials Engineering department, IISc Bengaluru. P.N. acknowledges Start-up grant from IISc, Infosys Young Researcher award, and DST-starting research grant SRG/2021/000285. R.R. acknowledges Science and Engineering Research Board for financial assistance (Grant Number: CRG/2021/000134). S.M. and P.N. acknowledge technical discussion with Dr. Subhajit Pal from School of Engineering and Materials Science, Queen Mary University of London, United Kingdom. We also acknowledge the help received from Dr. Gobinda Das Adhikary, Dr. Pramod Yadav, Shubham Kumar Parate, Monika Yadav and Deepak Sharma.

**Supplementary:**

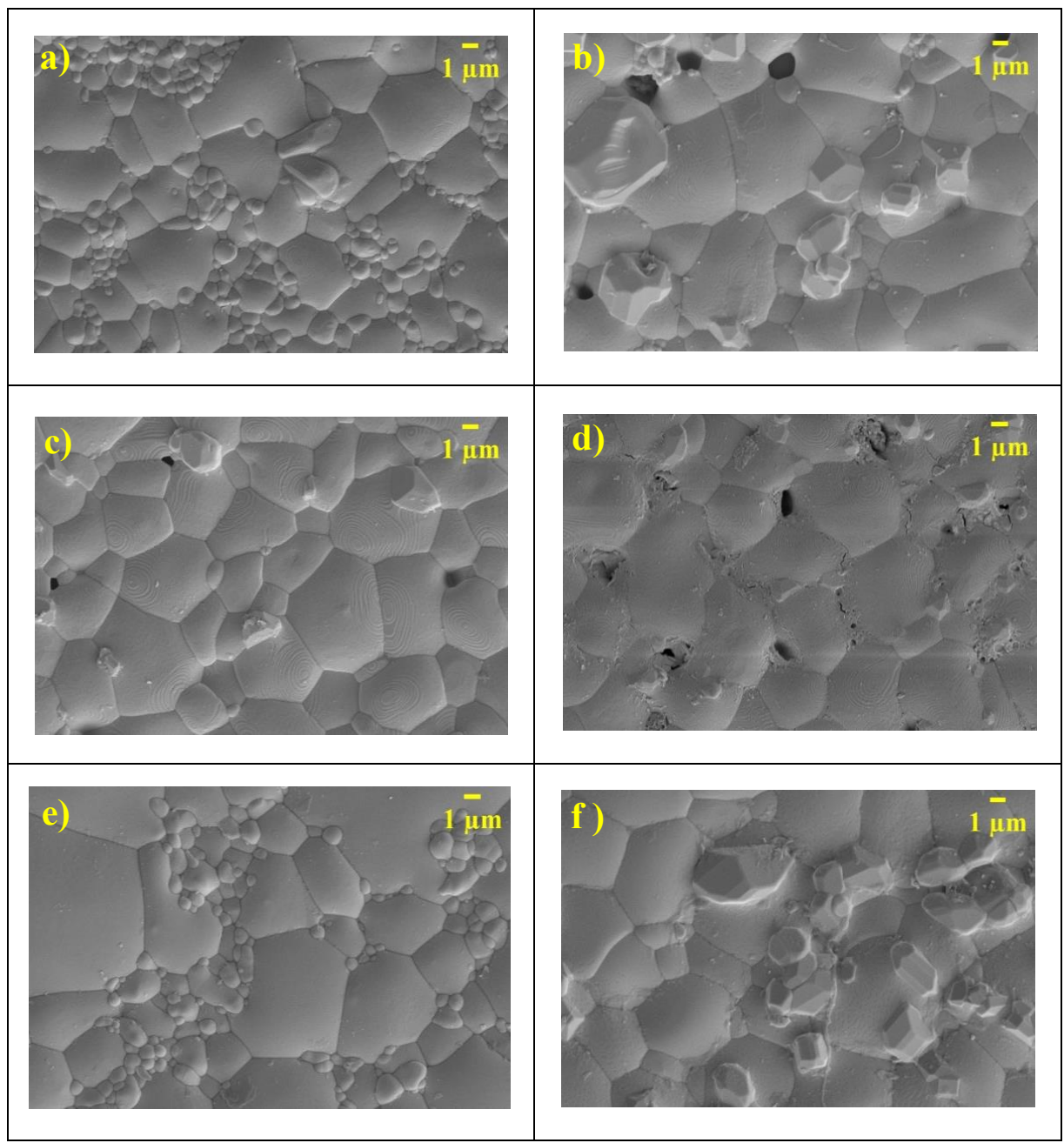

**Figure S1. SEM images of RE$_x$Ce$_{1-x}$O$_2$ (yRESC= y% RE substituted CeO$_2$).** a) 5GSC, b) 20GSC, c) 5LSC, d) 20LSC, e) 5YSC and f) 20YSC.

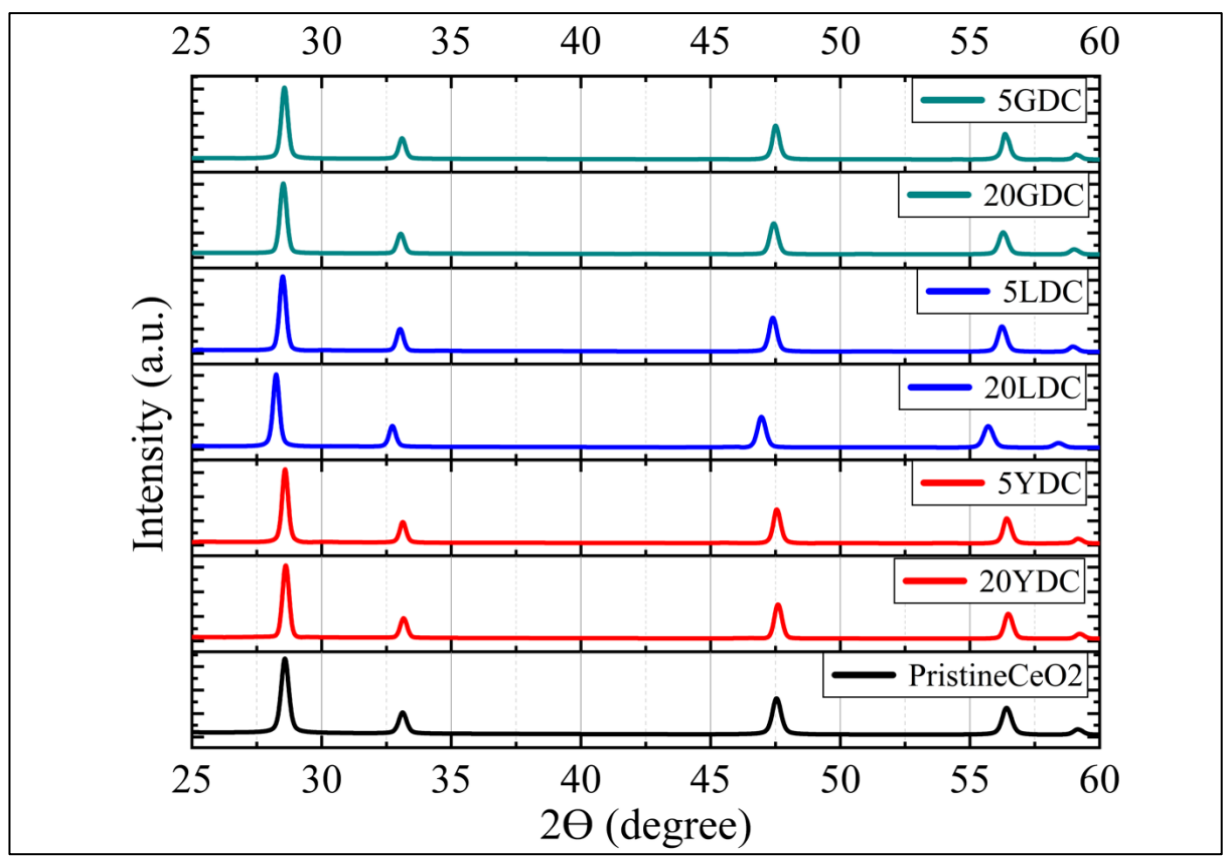

**Figure S2.** Rietveld refined powder XRD patterns of $RE_xCe_{1-x}O_2$ (yRESC= y% RE substituted $CeO_2$).

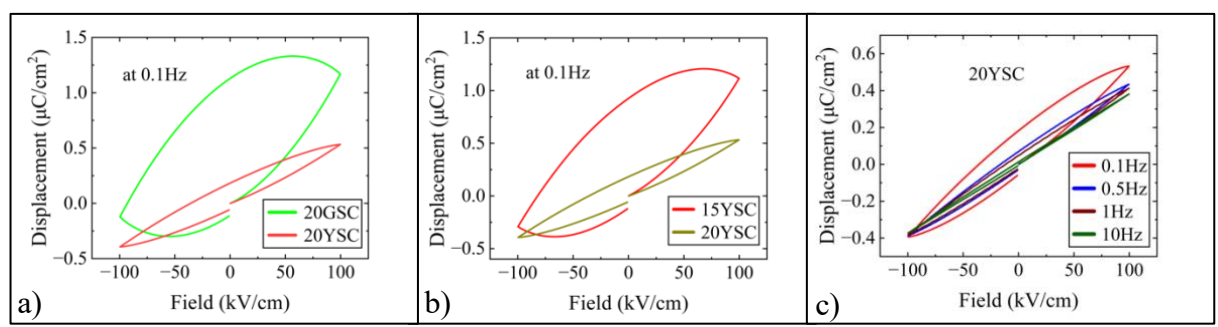

**Figure S3. Displacement v/s electric field plot** a) for $RE_{0.2}Ce_{0.8}O_2$ (RE= Gd, Y), b) for $Y_xCe_{1-x}O_2$ (x= 15, 20%) and c) for $Y_{0.2}Ce_{0.8}O_2$ at various frequencies.